\documentstyle[aps,prl,preprint]{revtex}

\begin{document}

\bibliographystyle{unsrt}
\tightenlines
 \newcommand{\um}[1]{\"{#1}}
  \newcommand{\uck}[1]{\o}
   \renewcommand{\Im}{{\protect\rm Im}}
    \renewcommand{\Re}{{\protect\rm Re}}
     \newcommand{\ket}[1]{\mbox{$|#1\protect\rangle$}}
      \newcommand{\bra}[1]{\mbox{$\protect\langle#1|$}}
       \newcommand{\proj}[1]{\mbox{$\ket{#1}\bra{#1}$}}
        \newcommand{\expect}[1]{\mbox{$\protect\langle #1 \protect\rangle$}}
         \newcommand{\inner}[2]{\mbox{$\protect\langle #1 | #2
         \protect\rangle$}}

\title{Experimental observation of nonclassical effects on single-photon
detection rates}  
\author{K. J. Resch, J. S. Lundeen and A. M. Steinberg}
\address{Department of Physics, University of Toronto\\
Toronto, ON M5S 1A7 CANADA}
\maketitle               

\begin{abstract}
It is often asserted that quantum effects can be observed in
coincidence detection rates or other correlations, but never in
the rate of single-photon detection.  We observe nonclassical 
interference in a singles rate, thanks to the intrinsic 
nonlinearity of photon counters.  This is due to a dependence
of the effective detection efficiency on the {\it quantum statistics} of
the
light beam.  Such measurements of detector response to photon
pairs promise to shed light on the microscopic aspects
of silicon photodetectors, and on general issues of 
quantum measurement and decoherence.
\end{abstract}

\newpage

Although quantum electrodynamics is one of the most well-established and
most accurately tested of physical theories, the vast majority of optical
physics still relies on the classical theory of the electromagnetic field.
This is because, as was recognized early in the development of quantum
optics\cite{glauber}, classical and quantum theories of light yield
precisely
the same predictions for a wide variety of phenomena.  In particular, the
two theories are equivalent for all calculations of intensities (i.e.,
mean photon number) and effects linear in the intensity; only by studying
higher-order correlations (e.g., Hanbury-Brown-Twiss-style coincidence
measurements\cite{hbt}) can one identify clear signatures of nonclassical
effects.  Over the past 20 years or so, many experiments have applied
coincidence-counting and autocorrelation techniques to access these
higher-order
correlation functions and demonstrate a variety of fascinating quantum
effects\cite{handbook}.  Since quantum and classical theories predict
identical intensity measurement outcomes,
such papers often begin by contrasting the featureless rate of single-photon
detection with the quantum phenomena observable {\it only} in coincidence  
rates.  Here, we show experimentally that even the rate of photodection
at a single detector can in fact be sensitive to quantum-statistical effects,
due to the frequently overlooked nonlinearity of typical photon counters
such as avalanche photodiodes.

The standard model of photodetection used in quantum optics dates back
to Glauber\cite{glauber}, and is appropriate for a low-efficiency
detector whose
photocurrent is, in the classical limit, proportional to the incident
intensity, or the photon number.  Real photon counters rely on extremely
nonlinear processes such as electron-hole-pair avalanches to amplify the
signal from a single photon to a detectable current.  As a result, nearly
all such devices (a rare exception, still not in widespread use, is
the Rockwell SSPM \cite{sspm,background}) are incapable of
distinguishing
between a single photon and two or more photons within a certain
(device-dependent) ``dead time''.  For an ideal, 100\%-efficient
detector this would imply, within the dead time, a photocount rate
proportional not to
$\expect{n} \equiv \expect{a^\dagger a}$, but rather to the step function
$\expect{\Theta(n-0.5)}$, which is equal to 1 for any photon number 
$n\ge1$. The violently nonlinear character of
this response
implies that even a single avalanche photodiode is sensitive not only
to the mean photon number, but also to higher-order terms.  If one restricts
oneself to considering photon numbers no higher than 2, for example, the
response function may be Taylor expanded as
$\frac{3}{2}\expect{n}-\frac{1}{2}\expect{n^2}$.  (This expansion can be
extended to include orders up to $n^j$ where
$j$ is the maximum number of photons considered in the subspace).  For a
real detector with a finite quantum efficiency, this model can be generalized.
However, no precise theoretical model for such a detector exists, nor has
the effect been experimentally studied prior to now.  Our na\"{\i}ve
model suggests that if the single-photon detection efficiency is $\eta$,
the probability of a detection event in the presence of two photons should
be $\eta + (1-\eta)\eta = 2\eta - \eta^2$.  This classical-minded
model neglects the possibility
of any cooperative effects between the photons, along with the possibility
of stimulated-emission processes in the detector.  Thus, measurements
of the ``two-photon detection probability'' should help one construct
accurate microscopic models of photon counters.  In particular, it may    
be possible to probe issues in quantum measurement theory in this way.
At very early times, the detection event is dominated by coherent processes
such as electron-hole pair-production.  As the avalanche process
amplifies this first pair, the system is coupled to a large number of
degrees of freedom and effective decoherence occurs.  Once coherence is
lost, our model should correctly describe the effect of
additional photons impinging on the detector.  If, however, a second
photon
arrives during the coherent phase of the detection process, it is reasonable
to expect a deviation from this prediction.

This experiment relies on a polarization-based version
\cite{Sergienko,shih}
of the two-photon interferometer orignally devised by Hong, Ou, and Mandel
\cite{hom}.  Our experimental setup is shown in figure 1.
Light from a mode--locked 800nm titanium:sapphire laser
is frequency doubled in a 0.2mm $\beta$--barium borate (BBO) 
crystal.  We separate the fundamental beam 
from the second harmonic with coloured glass filters (BG 39).
The second harmonic serves as the pump laser for the production of type-II 
degenerate down-conversion in a
second, 0.1mm, BBO crystal. 
The down-converted photon pairs exit the nonlinear 
crystal collinear with each other and the pump laser, and have orthogonal 
polarizations -- one photon's polarization is horizontal (H), and 
the other is vertical (V).  Due to BBO's birefringence, the V photon
trails the H photon by an average of 7 fs.  The pump laser
is separated from the downconverted
photons by using a fused-silica prism.
A 9mm thick quartz plate and a pair of translatable
birefringent quartz prisms are used to delay the photons 
relative to one another
inside the interferometer. The photon--pair polarization is rotated 
${45^{\circ}}$ 
by a half--wave plate and the pair goes through a cube polarizing
beamsplitter (PBS). The output
beams of the PBS travel through adjustable irises and are focused
onto 180$\mu$m avalanche photodiodes (APDs):  
(EG\&G model SPCM--AQR--13).  In this way, each photon has a 50\%
chance of reaching each detector; see figure 1.         

When the birefringent delay is set so that 
the photon wavepackets arrive at the PBS simultaneously, there is a drop in 
the coincidence rate.  The half--wave plate and the PBS serve to
measure the photons in the $\ket{\pm45^{\circ}}$ basis.  There are 
two possible Feynman paths
that lead to a coincidence event: photon 1 can be transmitted at the
PBS and photon 2 reflected, or vice versa. The amplitudes for 
these two processes
interfere destructively due to the different relative phases of $\ket{H}$
and $\ket{V}$ in the states
$\ket{\pm45^{\circ}}=(\ket{H}\pm\ket{V})/\sqrt{2}$, leading to
a coincidence
null provided the photons
arrive at the beamsplitter within their coherence length.  This
interference has no effect on the intensity of the light at either
detector.  If the photons
arrive at the beamsplitter outside each other's coherence length, then
they are
distinguishable, in principle, based on their arrival time at the
detectors.  In this case no interference is observed.

Theoretically, this coincidence dip can obtain a 
visibility of 100\%. However, for the purpose of this 
experiment, we obtained a higher 
photon collection efficiency and higher photon counting rates by opening
the irises in front of the detectors at the cost of lowering the
coincidence interference visibility.
In addition to using large irises, we employed a slightly imbalanced 1:1
telescope in the pump laser beam to again increase the collection
efficiency
\cite{brazil}.  

To our knowledge, there is no microscopic physical model for the response
of a APD when hit
by two photons at the same time. In order to give a
physically intuitive explanation for the origin and magnitude of an effect
on the singles rate, we analyze the Hong-Ou-Mandel interferometer using
our na\"{\i}ve model of detection. This model neglects the 
specifics of silicon APDs, and coherent effects between the multiple
electron-hole excitation pathways.
When the HOM interferometer is not balanced and there is no interference,
each photon-pair entering the device will lead to  0, 1, or 2 photons
reaching a given detector --
with probabilities 0.25, 0.5, and 0.25 respectively.  When
interference is occurring, half of the photon-pairs entering the device
will
impinge, as a pair, on the detector and the other half will leave through
the other exit port (for the ideal case of 100\% coincidence visibility).

Thus for $\eta=1$, the singles-counting rate drops from 75\% of the
pair-production rate to 50\%, a dip with 33\% visibility.  For a detector
with $\eta<1$ which obeys our model, the rate drops from 
$[\eta/2+(2\eta-\eta^{2})/4]$ to $[(2\eta-\eta^{2})/2]$, for a visibility
$V_{s}=\eta/(4-\eta)$.  For a realistic coincidence
visibility less than 100\%, we expect the
singles visibility
to fall off linearly with the coincidence visibility, $V_c$:
$V_s=V_c\eta/(4-\eta)$.

In order to investigate the singles effect and its dependence on the
collection 
efficiency, four runs were performed:
three with different
neutral density (ND) filters in front of one detector (Bob), and one 
with no ND filter.  By lowering the transmission of the ND filters, the
overall
collection efficiency can be lowered.  The 3 ND filters allowed for 80\%,
57\%, and 27\% transmission of light to give us a broad range of
efficiencies. The
irises in front of the detectors had 8mm diameters.  
The data were accumulated by scanning the quartz delay over the
interference 
dip many times.  An individual scan counted singles and coincidences at 280
evenly spaced delay settings for 1 second per point.  Typically 100 
such 5-minute scans were repeated for up to 20 hours, in
alternating directions to accumulate good statistics and minimize
systematic errors due to laser drift.  Over the 15
hours of data taking for the experiment with
no ND, Alice and Bob's
singles rates dropped by 10\%.

For the experiment
with no ND filter and with 8mm irises, the singles rates for Alice and Bob
were 13600/s and
13800/s respectively; the coincidence rate was 510/s.  By removing
the downconversion crystal, the backgrounds at Alice and
Bob were measured to
be 1415/s and 2006/s respectively.  Not including potential background
from the crystal itself, these background measurements are an
overestimate.  The crystal has imperfect transmission and
will attenuate both the pump laser and the on-axis background.  The
systematic error introduced by performing the measurement in this way 
could be as large as 20\%, which is much higher than the Poissonian
counting statistical error.  Based on the singles rates at the
beginning of the data collection, the background level
was measured to
be 9\% for Alice and 12\% for Bob.

The coincidence
and singles counts were binned and summed based on their delay position
to create the raw experimental data. The raw data 
contained fringes at the period of 800nm light which are due to a classical 
interference effect and have a visibility of approximately 1\%.  
These fringes are caused when the optic axis of the downconversion crystal
is not in the same plane as the optic axis in the quartz delay.  Since the 
period of these fringes is faster than the timescale for the 
HOM dip (2.6fs versus 20fs), these effects can be easily distinguished.  
The fringes were removed by averaging pairs of data points separated by
half the classical fringe period.  We fit the data to a Gaussian plus a
straight line, to
account for
residual effects of drift or nonuniform transmission through the quartz
prisms.

Prior to the data accumulation, the efficiency of detection was measured.  
This efficiency is the product of the
actual intrinsic quantum efficiency of the photodetector and the
path efficiency of the interferometer.  The path efficiency
includes losses due to finite sized irises, imperfect transmission
through optics, and the ND filters, but {\it not} the effect of the PBS.
In
order to determine the efficiency, $\eta_{A}$,
of one detector, the iris in front of the other detector is closed to
1-2mm \cite{klyshko,background}.  The efficiency is then
$\eta_{A}=2C_{AB}/(S_{B}-B_{B})$, where $C_{AB}$ is
the coincidence rate between Alice and Bob, $S_{B}$ is the singles rate at
Bob, $B_{B}$ is the background rate at Bob, and the factor of 2
compensates for the PBS.

Coincidence and singles rates are shown in figures 2 and 3 as a function
of the relative time delay, for the data run with no ND filters.  The data
have been binned and summed according to the relative time delay, and the
fringes have been removed by the
averaging method previously mentioned.  The coincidence visibility is 
$(39.39 \pm 0.05)\%$, and the raw singles visibility at Alice is
$(0.816 \pm 0.009)\%$, where the errors reflect only the statistical
fitting uncertainties. Correcting for the $(9.4\pm1.9)\%$
background that we measured at Alice the corrected visibility is
$(0.90\pm0.01)\%$.  The efficiency was measured to be $(8.4\pm1.1\%)$,
the large uncertainty being due to the sensitivity to background when
small irises are used.  Our model predicts a
visibility of $(0.85\pm0.11)\%$
based on the uncertainty in the efficiency measurement. The centres and
widths of the coincidence dip and
Alice's singles rate dip agree to within just over one standard deviation.

Since the expected singles visibility scales linearly with the
coincidence visibility, figure 4 shows their ratio versus efficiency. Both
the singles visibility and the efficiency are corrected for background.
There are two sets of error bars plotted on each data point.  The smaller
set of error bars represents the statistical errors only.  The larger set
of error bars includes the systematic error due to the
uncertainty in the background.  
The singles visibility at Bob is linearly proportional to the efficiency.
The four data points for Bob
fall on a straight line (dashed line) with a zero y-intercept with a
$\chi^2$ of 5.6
based solely on the statistical errors.  
The slope differs from that of our theoretical curve (solid line), but by
an amount
attributable to systematic uncertainty in background.
We have made preliminary measurements that suggest that 
we are missing much of the background contribution to the singles 
rates at both of our detectors.  This is due to fluorescence
background which is created in 
the downconversion crystal or in its anti--reflection coatings, and is
difficult to to isolate from the signal itself.  Refinement of
background measurement techniques are an important subject for future
work. It
should also be mentioned that since the background measurement is
incorporated into the values for the efficiency and for the singles
visibility, a higher background will increase both values, but by
different amounts.

We have demonstrated for the first time a quantum effect on the counting
rate of a single-photon counter, showing that its effective
efficiency is a function of the quantum statistics of the incident beam.
A simple model for this effect is roughly consistent
with the data; however, it will be necessary to refine our estimates of
various background sources in order to further pursue quantitative tests
of this model.

By adjusting the correlation properties of the incident beam, we are
able to probe the response of Silicon avalanche photodiodes to photon
pairs (as compared with single photons)
on timescales of the order of 5 fs, and for delays up to 60 fs.
Since energy and momentum relaxation timescales in Silicon are on the
order of 100 fs, and the incident photons may be prepared in identical
quantum states (with focal spots on the order of 180$\mu$m), a
quantum-mechanical description of the microscopic
detection process might be expected to be necessary in this regime.    
At present, the dependence of this effect on detector efficiency agrees
qualitatively with a simple classical model, and the agreement can be
made quantitative when a 20\% systematic uncertainty in background is
included.  For pairs of photons
separated by delays ranging from 20 fs to 60 fs, well outside the
interference region, no systematic
variation in the counting rate was observed at the 0.1\% level.  
Over the 10-fs width of the quantum
interference pattern, the shapes of the singles and coincidence pattern
are the same to within 2.5\%. Thus, quantum corrections to our model of
the detector response to photon pairs appear to
be quite small in the regime probed so far.  By better characterizing the
backgrounds and by extending these studies to other temporal or spatial
scales or other detectors, the type of system presented here should
provide a novel tool for studying decoherence and the measurement problem
at ultrafast timescales in a variety of physical systems.    

We would like to thank Magali Davenet and Chris Dimas
for their assistance with this experiment, and
Sasha Sergienko, John Sipe, and Henry van Driel for extremely valuable
discussions. 
We would also like to acknowledge NSERC, Photonics Research 
Ontario (PRO), the Canadian Foundation for Innovation and the Walter C.
Sumner foundation for financial support.

\newpage

Figure 1:  Experimental setup for polarization-based two-photon
interferometer: BG 39,
colour glass filter; $\frac{\lambda}{2}$, half-wave plate; PBS, polarizing
beamsplitter; Alice/Bob, single photon counting modules.  
Different transmission ND filters were used for each data
acccumulation run.

Figure 2:  Typical plot of coincidence rates versus the relative time
delay.
The Gaussian fit yields a visibility of $(39.39\pm0.05)\%$.

Figure 3:  Typical plot of the singles rate versus the relative time
delay. The
gaussian
fit yields a visibility of $(0.816\pm0.009)\%$.  When background is
included,
the visibility becomes $(0.90\pm0.01)\%$, based on statistical errors
only.
With the measured efficiency for Alice of $(8.4\pm1.1)\%$, including the
systematic error, our model predicts a visibility of $(0.84\pm0.11)\%$.

Figure 4:  Plot of the ratio of singles visibility to coincidence
visibility versus the collection efficiency.  The solid square is a data point
from Alice's singles visibility, and the open circles are from Bob's
singles visibility.  The solid line is the prediction of our na\"{\i}ve
model.  The dashed line is a linear fit with a zero y-intercept to the
data for Bob based on statistical errors only. The data are shown with
two sets of error bars. The smaller error
bars represent statistical uncertainties only, and the larger set include
both the statistical and the systematic errors.

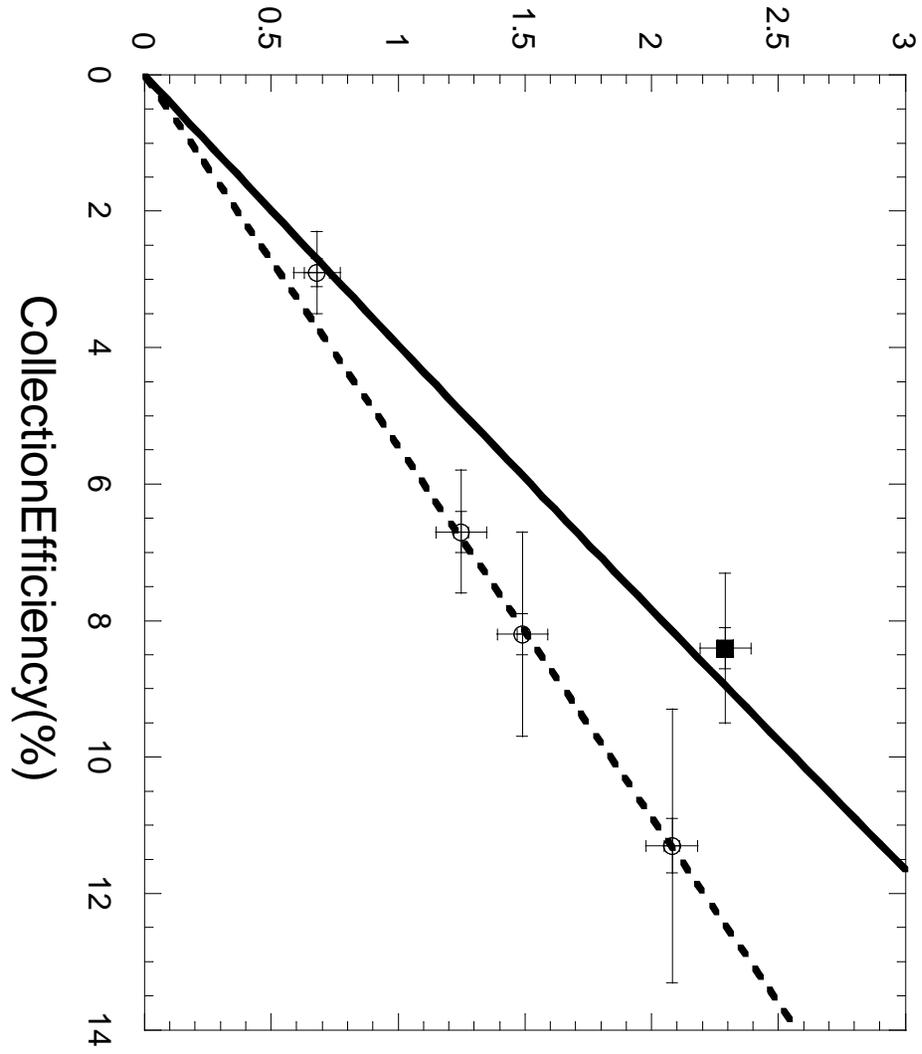
\end{document}